\documentclass[12pt,preprint]{aastex}

\shorttitle{Photon emissivity of the electrosphere}
\shortauthors{Harko \& Cheng}
\usepackage{subfigure}
\begin{document}
\title{Photon emissivity of the electrosphere of bare strange stars}
\author{T.~Harko$^1$ and K.~S.~Cheng$^2$}
\affil{Department of Physics, The University of Hong Kong,
Pokfulam Road, Hong Kong, Hong Kong SAR, P. R. China}
\email{$^1$harko@hkucc.hku.hk, $^2$hrspksc@hkucc.hku.hk}

\begin{abstract}
We consider the spectrum, emissivity and flux of the
electromagnetic radiation emitted by the thin electron layer (the
electrosphere) at the surface of a bare strange star. In
particular, we carefully consider the effect of the multiple and
uncorrelated scattering on the radiation spectrum (the
Landau-Pomeranchuk-Migdal effect), together with the effect of the
strong electric field at the surface of the star. The presence of
the electric field strongly influences the radiation spectrum
emitted by the electrosphere. All the radiation properties of the
electrons in the electrosphere essentially depend on the value of
the electric potential at the quark star surface. The effect of
the multiple scattering, which strongly suppresses radiation
emission, is important only for the dense layer of the
electrosphere situated near the star's surface and only for high
values of the surface electric potential of the star. Hence a
typical bremsstrahlung radiation spectrum, which could extend to
very low frequencies, could be one of the main observational
signatures even for low temperature quark stars.
\end{abstract}


\section{Introduction}

The quark structure of the nucleons, suggested by quantum
chromodynamics, indicates the possibility of a hadron-quark phase
transition at high densities and/or temperatures, as suggested by
\citep{It70,Bo71,Te79,Wi84}. Theories of the strong interaction,
like, for example, the quark bag models, suppose that breaking of
physical vacuum takes place inside hadrons. If the hypothesis of
the quark matter is true, then some of neutron stars could
actually be strange stars, built entirely of strange matter
\citep{Al86,Ha86}. For a general review of strange star
properties, see \citet{Ch98}.

There are several proposed mechanisms for the formation of quark
stars. Quark stars are expected to form during the collapse of the
core of a massive star, after the supernova explosion, as a result
of a first or second order phase transition, resulting in
deconfined quark matter \citep{Da}. The proto-neutron star core or
the neutron star core is a favorable environment for the
conversion of ordinary matter to strange quark matter
\citep{ChDa}. Another possibility is that some neutron stars in
low-mass X-ray binaries can accrete sufficient mass to undergo a
phase transition to become strange stars \citep{Ch96}. This
mechanism has also been proposed by \citet{Ch98a} as a source of
radiation emission for cosmological $\gamma $-ray bursts. Quark
stars can also be formed during the rapid spin-down of magnetars,
astrophysical objects with extremely high magnetic fields
\citep{HaChTa04}.

Based on numerical integration of the general relativistic
hydrostatic equilibrium equations a complete description of the
basic astrophysical properties (mass, radius, eccentricity,
Keplerian frequency etc.) of both static and rotating strange
stars can be obtained \citep{Wi84,Ha86,Go00,De98,ChHa,HaCh02}.
Rotational properties can discriminate between neutron and quark
stars. Strange stars can reach much shorter periods than neutron
stars, of the order of $0.5$ ms \citep{Ch98}. $r$-mode
instabilities in rapidly rotating strange stars lead to specific
signatures in the evolution of pulsars with periods below $2.5$
ms. If strange matter is absolutely stable, pulsars would be
expected to consist of quark matter. Some data on pulsar
properties are consistent with this assumption \citep{Ma00}.


The structure of a realistic strange star is very complicated, but its basic
properties can be described as follows \citep{Al86}. Beta-equilibrated
strange quark - star matter consists of an approximately equal mixture of up
$u$, down $d$ and strange $s$ quarks, with a slight deficit of the latter.
The Fermi gas of $3A$ quarks constitutes a single color-singlet baryon with
baryon number $A$. This structure of the quarks leads to a net positive
charge inside the star. Since stars in their lowest energy state are
supposed to be charge neutral, electrons must balance the net positive quark
charge in strange matter stars. The electrons, being bounded to the quark
matter by the electromagnetic interaction and not by the strong force, are
able to move freely across the quark surface, but clearly cannot move to
infinity because of the electrostatic attraction of quarks. For hot stars the electron
distribution could extend up to $\sim 10^{3}$ fm above the quark surface \citep{ChHa03}.

Photon emissivity is the basic parameter for determining
macroscopic properties of stellar type objects. \citet{Al86} have
shown that, because of very high plasma frequency $\omega _p$ near
the strange matter edge, photon emissivity of strange matter is
very low. Only photons produced just below the surface, in a quark
layer of approximately $5$ fm, and with momenta pointing outwards,
can leave strange matter. For temperatures $T<<E_{p}/\omega $,
where $E_{p}\approx 23$ MeV is the characteristic transverse
plasmon cutoff energy, the equilibrium photon emissivity of
strange matter is negligible small, as compared to the
black body one. The spectrum of equilibrium photons is very hard, with $%
\hbar \omega >20$ MeV.

The bremsstrahlung emissivity of quark matter has
been estimated by \citet{Chmaj}. They have found
that the surface radiation is about four orders of magnitude weaker than the
equilibrium black body radiation, even they both have the same temperature
dependence. The problem of the soft photon emissivity of quark matter at the surface
of strange stars has been reconsidered in \citet{ChHa03}. By taking into account the effect of
the interference of amplitudes of nearby interactions in a dense media (the
Landau-Pomeranchuk-Migdal effect) and the absorption of the radiation in the
external electron layer, the emissivity of the quark matter can be six
orders of magnitude lower than the equilibrium black body radiation.

The Coulomb barrier at the quark surface of a hot strange star may also be a
powerful source of $e^{+}e^{-}$ pairs, which are created in the extremely
strong electric field of the barrier. At surface temperatures of around $%
10^{11}$ K, the luminosity of the outflowing plasma may be of the order $%
\sim 10^{51}$ ergs$^{-1}$ \citep{Us98,Us98a}. Moreover, as shown by \citet{PaUs02}, for about one day for normal quark matter and for up to a
hundred years for superconducting quark matter, the thermal luminosity from
the star surface, due to both photon emission and $e^{+}e^{-}$ pair
production may be orders of magnitude higher than the Eddington limit.

The existence of another major source of energy from a bare
strange star, namely, the bremsstrahlung photon emission from the
high-density electron layer at the surface of the star, has been
pointed out by \citet{Ja04a,Ja04b}. The electron layer could be an
important source of photons, which, due to the extremely low
emissivity of quark-gluon plasma, could be the main observational
signature of strange stars. The extreme-relativistic electrons of
the electron layer undergo several types of collisions. The most
important processes are the electron-electron elastic scattering,
electron-photon scattering (Compton scattering), and
electron-electron bremsstrahlung (together with its inverse). Of
these three, the first two are assumed, on the average, to cause
no net change of the gas, i.e., the electron and photon
distributions remain unchanged. Hence the most important effect
leading to the creation and subsequent loss of radiation in the
electrosphere of the quark stars is the bremsstrahlung emission.

Since the electrons in the electrosphere of the bare strange stars
have a high density, the effect of the dense medium plays an
important role on the electromagnetic emission of the thin
electron layer. The intensity of radiation from high density
systems can be significantly reduced by the so-called
Landau-Pomeranchuk-Migdal (LPM) effect
\citep{LaPo53,Mi56,An95,Kl99,Ha03,ChHa03}, which takes into
account the effect of multiple collisions on the intensity of the
emitted radiation. The influence of the LPM effect for the
bremsstrahlung emisivity of the electrosphere has been discussed
in \citet{Ja04b} and it has been shown that the effect of the
dense medium can reduce the emissivity by two orders of magnitude.

It is the purpose of the present paper to consider a systematic
analysis of the LPM effect in the electrosphere of the bare quark
stars. In particular, we would like to point out the effect of the
electric field at the quark star surface on the electromagnetic
radiation from the electrosphere. The presence of the electric
field considerably reduces the magnitude of the LPM effect (for
example, the critical LPM frequency is reduced, due to the
electric field, from $1$ GeV to a few MeV).

The radiation properties of the electrosphere essentially depend
on the value of the electric potential at the quark star surface.
For high values of the electric potential (of the order of $16-20$
MeV or higher), the electrosphere can be considered, from the
point of view of the LPM effect, as a thick medium, and radiation
is strongly suppressed in almost the entire volume of the
electrosphere. On the other hand, for small values of the surface
electrostatic potential of the quark star the electrosphere
becomes thin and the suppression effect can be ignored.


The present paper is organized as follows. In Section II we review
the basic formalism and the main results describing the radiation
spectrum of the individual electrons by taking into account the
LPM effect. The emissivity and energy flux of the electrosphere is
obtained in Section III. We discuss and conclude our results in
Section IV. Throughout this paper we use the natural system of
units, with $c=\hbar =k_B=1$.

\section{Photon emissivity and electromagnetic radiation energy flux from
the electrosphere of the quark stars}

In the present Section we shall review some of the basic
definitions and physical processes related to the electromagnetic
radiation emission the of electrons in the dense electron layer at
the surface of a quark star. In particular, we shall describe in
some details the mechanisms of electromagnetic radiation
suppression due to the Landau-Pomeranchuk-Migdal effect in the
case of electrons in a dense medium \citep{LaPo53,Mi56}, with and
without the presence of an external electric field. This effect
plays an essential role in the calculation of the radiation
emission from the electrosphere of the quark star.

\subsection{Bremsstrahlung emissivity and energy flux of the
electrosphere}

The emission of a photon in the electrosphere of a quark star is
the result of the scattering of an electron from a state $\vec{p}_{1}$ to a state $\vec{%
p}_{1}^{\prime }$ by collision with another particle of momentum
$\vec{p}_{2} $. In order to calculate the photon emissivity due to
electron collisions in the electrosphere of quark stars we
consider the rate of collisions $dN$ in which a soft photon of
energy $\omega $ is emitted. According to the general principles
of quantum electrodynamics the rate of collisions is given by
\citep{La}
\begin{equation}
dN=\frac{1}{V}\int dWf_{1}\left( \epsilon _{1}\right) f_{2}\left(
\epsilon _{2}\right) \times \left[ 1-f_{1}^{\prime }\left(
\epsilon _{1}^{\prime }\right) \right] \left[ 1-f_{2}^{\prime
}\left( \epsilon _{2}^{\prime }\right) \right] ,
\end{equation}
where
\begin{equation}
dW=V\left( 2\pi \right) ^{4}\delta ^{(4)}\left(
p_{1}+p_{2}-p_{1}^{\prime }-p_{2}^{\prime }-k\right) \left|
M\right| ^{2}\frac{d^{3}k}{\left( 2\pi
\right) ^{3}2\omega }\prod_{i=1,2,1^{\prime },2^{\prime }}\frac{d^{3}p_{i}}{%
\left( 2\pi \right) ^{3}2\epsilon _{i}},
\end{equation}
is the scattering rate. We have denoted by $\left( \epsilon _{i},\vec{p}%
_{i}\right) $ the i-th electron energy and momentum, where
$\epsilon
_{i}=\left( p_{i}^{2}+m^{2}\right) ^{1/2}$, and by $\left( \omega ,\vec{k}%
\right) $ the photon energy and momentum.  $f(\left[ \epsilon (p)-\mu _{e}%
\right] =1/\left( \exp \left[ \frac{\epsilon (p)-\mu
_{e}}{T}\right]
+1\right) $ is the Fermi-Dirac distribution function for the i-th electron %
\citep{Ch68} and $M$ is the scattering amplitude. In order to
obtain an explicit expression for the scattering rate we adopt the method developed in %
\citet{La}, by assuming that the photon emission process is
quasi-classical. Consequently, for soft photon emission $dW$ can
be written in a factorized form, $dW=dW_{0}dW_{\gamma }$, with
$dW_{0}$ being the elastic electron-electron scattering rate and
$dW_{\gamma }$ the probability of
emission of a photon of energy $\omega $ by any of the scattered electrons. $%
dW_{0}$ can be obtained from
\begin{equation}
dW_{0}=V\left( 2\pi \right) ^{4}\delta ^{(4)}\left(
p_{1}+p_{2}-p_{1^{\prime }}-p_{2^{\prime }}\right) \left|
M_{0}\right| ^{2}\prod_{i=1,2,1^{\prime },2^{\prime
}}\frac{d^{3}p_{i}}{\left( 2\pi \right) ^{3}2\epsilon _{i}},
\end{equation}
where $M_{0}$ is the electron-electron scattering amplitude. The
reaction rate can be expressed in a more convenient form in terms
of the cross section $\sigma $ of the reaction.

In terms of the cross section $\sigma $ of the emission process the rate of collisions in the electron plasma in which a photon of energy $%
\omega <<\epsilon $ is emitted can be obtained as \citep{Ch68}
\begin{equation}
dN=g_{1}g_{2}f_{1}\left( \vec{p}_{1}\right) f_{2}\left(
\vec{p}_{2}\right)
\sigma \left( \theta ,\phi \right) \left| \vec{v}_{1}-\vec{v}_{2}\right| %
\left[ 1-f_{1}\left( \vec{p}_{1}^{\prime }\right) \right] \left[
1-f_{2}\left( \vec{p}_{2}^{\prime }\right) \right]
\prod_{i=1,2,1^{\prime },2^{\prime }}d^{3}\vec{p}_{i}d\Omega
\left( \theta ,\phi \right) ,
\end{equation}
where $d\Omega \left( \theta ,\phi \right) $ is the solid angle
element in the direction $\left( \theta ,\phi \right) $,
$g_{1},g_{2}$ are the statistical weights and $\vec{v}_{1}$,
$\vec{v}_{2}$ are the velocities of the interacting particles. The
invariant relative velocity $v_{r}=\left|
\vec{v}_{1}-\vec{v}_{2}\right| $ is defined by $v_{r}=\sqrt{\left[
\left( p_{1}\cdot p_{2}\right) ^{2}-m_{e}^{4}\right] /\left(
p_{1}\cdot p_{2}\right) ^{2}}$, where $p_{1}$ and $p_{2}$ are the
energy-momentum four-vectors of the two electrons \citep{La75}.
The metric has been chosen so that the product of two four-vectors
$a^{\mu }$ and $b^{\mu }$ is defined by $a\cdot b$
$=a_{0}b_{0}-\vec{a}\cdot \vec{b}$. The energy $\epsilon $ of the
electron in the center of mass system is related to the
$4$-momenta in
an arbitrary frame by $\epsilon =\left[ \left( p_{1}\cdot p_{2}+1\right) /2%
\right] ^{1/2}$ \citep{La75}. Therefore in order to calculate the
reaction rate for the photon emission process we need to know the
cross section of this process.

Let $d\sigma _{0}$ be the cross-section for a given process of
scattering of charged particles, which may be accompanied by the
emission of a certain number of photons. For example, $d\sigma
_{0}$ could refer to the scattering of an electron by another
electron, with the possible emission of hard photons. Together
with this process one could consider another process which differs
from it only in that one extra photon is emitted. In this case the
total cross-section $d\sigma $ can be represented as a product of
two independent factors, the cross-section $d\sigma _{0}$ and the probability $%
dW_{\gamma }$ of emission of a single photon in the collision
\citep{La}. The emission of a soft photon is a quasi-classical
process. The probability of emission is the same as the
classically calculated number of quanta emitted in the collision,
that is the same as the classical intensity (total energy) of
emission $dI$, divided by the frequency of the radiation $\omega $
\citep{La}.

The probability $dW_{\gamma }$ of emitting a photon of energy
$\omega $ by any of the scattered electrons is $dW_{\gamma
}=dI/\omega $. Therefore the total cross section for the emission
of soft photons is given by \citep{La}
\begin{equation}
d\sigma =d\sigma _{0}\frac{dI}{\omega }.  \label{cross}
\end{equation}

To calculate the total cross section for a photon emission we need to know $%
d\sigma _{0}$, describing the elastic scattering of electrons. We
shall assume that $d\sigma _{0}$ is equal to the electron-electron
elastic scattering cross-section $d\sigma _{el}$, given, in the
center of mass
system, in which the magnitudes of the momentum $\vec{p}$ and energy $%
\epsilon $ of the electrons are unchanged in the scattering, by
\citep{La}
\begin{equation}
d\sigma _{el}=r_{e}^{2}\frac{m_{e}^{2}\left( \epsilon
^{2}+\vec{p}^{
2}\right) ^{2}}{4\vec{p}^{4}\epsilon ^{2}}\left[ \frac{4}{\sin ^{4}\theta }-%
\frac{3}{\sin ^{2}\theta }+\left( \frac{\vec{p}^{2}}{\epsilon ^{2}+\vec{p}%
^{2}}\right) ^{2}\left( 1+\frac{4}{\sin ^{2}\theta }\right)
\right] d\Omega .
\end{equation}

By assuming that the energies of the electrons in the
electrosphere are ultra-relativistic, $\vec{p}^{2}=\epsilon ^{2}$,
we obtain from Eq. (\ref {cross}) the limiting form $d\sigma
_{el}=\left( r_{e}^{2}m_{e}^{2}/\epsilon
^{2}\right) \left( 3+\cos ^{2}\theta \right) ^{2}d\Omega /4\sin ^{4}\theta $%
. Therefore the total cross section for the emission of a photon
can be written as
\begin{equation}
\sigma =\int d\sigma _{el}\left\langle \frac{dI}{d\omega
}\right\rangle \frac{d\omega }{\omega },
\end{equation}
where the factor $\left\langle dI/d\omega \right\rangle$ describes
the intensity of the radiation emitted by the electron in multiple
collisions.

We assume that the electrons form a degenerate Fermi gas, with the
particle number density $n$ given by \citep{Ch68}
\begin{equation}
n=\frac{g}{2\pi ^{2}}\int_{0}^{\infty }f\left[ \epsilon (p)-\mu
_{e}\right] p^{2}dp=\frac{g}{2\pi ^{2}}\int_{0}^{\infty
}\frac{p^{2}dp}{\exp \left[ \frac{\epsilon (p)-\mu _{e}}{T}\right]
+1}.
\end{equation}

The emissivity per unit volume of the electrosphere can be
calculated by using the definition
\begin{equation}
\varepsilon _{\gamma }^{(ee)}(z,T)=\frac{dE_{\gamma
}^{ee}}{dtdV}=\int \omega dN.
\end{equation}

The energy flux from the electrosphere of a strange star, coming
out from a thin surface layer of thickness $dz$, is $F_{\gamma
}^{(ee)}=\varepsilon _{\gamma }^{(ee)}dz/\pi $. Taking into
account the contribution of all layers we find
\begin{equation}
F_{\gamma }^{(ee)}(T)=\frac{1}{\pi }\int_{0}^{\infty }\varepsilon
_{\gamma }^{(ee)}\left( z,T\right) dz.
\end{equation}

\subsection{Radiation intensity of charged particles in dense
media}

As a first step in obtaining the photon emissivity of the
electrosphere of the bare strange stars we have to obtain the
classical radiation intensity distribution $<dI/d\omega >$,
emitted by electrons moving in a dense medium in which many
inter-particle collisions occur. Since in the electrosphere a very
strong electric field is also present, the influence of this field
on the radiation process must also be taken into account.

In the early 1950s \citet{LaPo53} and \citet{Mi56}predicted that
the radiation from highly relativistic particles in dense media is
suppressed due to interference between amplitudes of nearby
interactions (for detailed presentations of the LPM effect see the
monograph by \citet{AkSh96} and the review article by
\citet{Kl99}). The suppression effect is of quantum mechanic
nature and has its roots in the uncertainty principle. The
kinematics of the bremsstrahlung requires that the longitudinal
momentum transfer between the interacting particles must be small;
the uncertainty principle then requires that the interaction must
occur over a large longitudinal distance scale (formation zone).
If the electron Coulomb scatters while traversing this zone, the
bremsstrahlung amplitude from before and after scattering can
interfere, thus reducing the amplitude for bremsstrahlung photon
emission (\citet{AkSh96} and references therein). The results of
the quantitative measurements of the bremsstrahlung suppression
due to the LPM effect for electrons with energies of $8$ and $25$
GeV traversing thin gold and carbon targets have been presented in
\citet{An95}. The suppression of bremsstrahlung predicted by the
LPM theory is correct to within $5\%$.

The main parameters defining radiation processes in a medium are
the coherence length $l_c$, the mean free path of a fast particle
in matter $l_{e}$, the radiation length $X_{0}$ and the thickness
of the target $L$ (\citet{AkSh96} and references therein). The
coherence length is the distance along the particle momentum where
interference effects during the radiation process are significant.
$l_c$ rapidly grows with an increase of the particle energy and at
high energies it can be macroscopic. If the energy $\omega $ of
the bremsstrahlung photon produced by an ultra-relativistic
particle of energy $E$ satisfies the condition $\omega <<\epsilon
$, then the average angle $\theta $ between the incident particle
and the photon is small, $\theta _c\approx mc^2/\epsilon $. The
average angle between the scattered particles is smaller still.
Neglecting these angles, the longitudinal momentum transfer
between the interacting particles is $p_{\left| {}\right| }\simeq
\omega /2\gamma ^{2}c$, where $\gamma =\epsilon /mc^{2}$. The
uncertainty principle then requires that the spatial position of
the bremsstrahlung process has a longitudinal uncertainty of
$l_{c}=\hbar /p_{\left| {}\right| }\approx 2\hbar c\gamma
^{2}/\omega $. The coherence length rapidly grows with an increase
of the particle energy, and at high energies can be macroscopic.
In alternate language, the particle and photon slowly split apart
over the distance $l_{c} $. In a sufficiently dense medium the
mean free path of the incident particle is much smaller than
$l_{c}$, so a particle will interact while traversing the region
$l_{c}$. Bremsstrahlung is suppressed when the mean square
multiple scattering angle over the distance $l_c$,
\begin{equation}\label{th}
\theta _{s}^{2}=\left( \frac{\epsilon _{s}}{\epsilon }\right)
^{2}\frac{l_{c}}{X_{0}},
\end{equation}
is greater than or equal to $\theta _{k}^{2}$ (\citet{AkSh96,Kl99}
and references therein). Here
\begin{equation}\label{x0}
X_{0}=\left[ 4n\alpha r_{e}^{2}Z^{2}\ln \left( 184Z^{-1/3}\right)
\right] ^{-1},
\end{equation}
is the radiation length and
\begin{equation}
\epsilon _{s}=mc^{2}\sqrt{\frac{4\pi }{\alpha }}.
\end{equation}

In Eq. (\ref{x0}) $r_{e}=e^{2}/mc^{2}$ is the classical radius of
the particle (the electron). The effect of multiple suppression on
the total radiation emission of the charged particle, $dI/d\omega
$, can be obtained, for $\omega <<\epsilon _{s}^{2}c/\epsilon
^{2}X_{0}$, in the form (\citet{Kl99} and references therein)
\begin{equation}\label{est}
\frac{dI}{d\omega }=\sqrt{\frac{2\pi }{3}}\frac{Z^{2}e^{2}m^{2}c^{3}}{\epsilon _{s}\epsilon }%
\sqrt{\frac{\omega X_{0}}{c}}.
\end{equation}

Thus, multiple scattering results in a decrease of the coherence
length, leading, in turn, to radiation suppression. The effect
appears in the case when the mean-square angle of the multiple
scattering within the coherence length $\left\langle \theta
^{2}\right\rangle $ exceeds the characteristic squared angle of
the relativistic electron radiation $\theta ^{2}\sim 1/\gamma
^{2}$ \citep{LaPo53,Kl99}. The suppression happens when the
frequency of the radiation satisfies the condition
\begin{equation}
\omega <\omega _{LPM}=\frac{\epsilon _{s}^{2}\epsilon
}{m^{4}c^{7}X_{0}}.
\end{equation}

The electromagnetic radiation of charged particles in external
fields plays an essential role in astrophysical processes, the
most important cases being the synchrotron radiation in a constant
magnetic field or the unified synchro-curvature radiation of
electrons in constant electric and magnetic fields
\citep{HaCh02b}. However, the radiation of electrons in pure
electric fields $\vec{E}$ has been less studied.

The intensity of the radiation of charged particles in dense media
in the presence of an external electric field and with the LPM
effects included was obtained by \citet{BaTi86} and later on by
\citet{TBHa97}. In the present paper we shall closely follow the
presentation and the derivation of \citet{BaTi86}. Due to the
presence of the electric field, there is a decrease of the
influence of multiple scattering on the process of radiation.

This important result can be understood qualitatively as follows
\citep{BaTi86}. In a dense media the condition for the suppression
of the radiation due to multiple scattering is $\left\langle
\theta ^{2}(l_c)\right\rangle \geq \theta _{\gamma }^{2}$, where
$\left\langle \theta ^{2}\right\rangle $ is given by Eq. (\ref
{th}) and $l_{c}\approx 2\gamma ^{2}/\omega \approx 2\epsilon
^2/m^2\omega $ is the coherence length. In obtaining this relation
we have assumed that the scattering angles are randomly oriented
in the plane perpendicular to the direction of motion of the
particle.

For sufficiently high electron energies we can approximate $\omega
\approx \epsilon $, thus obtaining $l_c\approx 2\epsilon
/m^{2}$ and $\left\langle \theta ^{2}(l_c)\right\rangle \propto \epsilon ^{-1}$%
. Since $\theta _{\gamma }^{2}\propto \epsilon ^{-2}$, at
sufficiently high energies the condition $\left\langle \theta
^{2}(l_c)\right\rangle \geq \theta _{\gamma }^{2}$ will always be
satisfied.

However, this situation radically changes in the case of a
particle in an external electric field. Radiation in the electric
field of intensity $E$ is characterized by the parameter
\begin{equation}
\zeta =\frac{eE\epsilon }{m^{3}}=\frac{\gamma E}{E_{cr}},
\end{equation}
where $E_{cr}=m^2/e$ is the critical electric field \citep{La}.
For electrons, the value of $E_{cr}$ is $E_{cr}=1.3\times 10^{16}$
V/cm. If $\zeta \geq 1$, the energy of the radiated  gamma rays is
$\omega \sim \zeta \epsilon /\left( 1+\zeta \right) \sim \epsilon
$, and the radiation process is essentially quantum in its nature.

For $\zeta>>1$ the characteristic angle of the radiation $\theta
_{\gamma }$ is modified to $\theta _{E}\propto \theta _{\gamma
}\zeta ^{1/3}$ and the coherence length becomes $l_{E}\propto
\omega /m\zeta ^{2/3}$ \citep{BaTi86}. Multiple scattering will
greatly change the pattern of formation of radiation for
$\left\langle
\theta ^{2}(l_{E})\right\rangle \geq \theta _{E}^{2}$ . Since $\zeta \sim \epsilon $%
, we have $\left\langle \theta ^{2}(l_{E})\right\rangle \propto
\epsilon ^{-5/3}$ and $\theta _{E}^{2}\propto \epsilon ^{-4/3}$,
i.e., in contrast to the case of a dense medium without an
external field, the left hand side of the inequality falls off
with the increase of the energy $\epsilon $ faster than the right
side. The suppression of the LPM effect in the presence of the
electric field can be interpreted as an effective shortening of
the coherence length because in a strong electric field the
$1/\gamma $ emission zone shortens- the emitting particle is
deviated out of the emission cone by the field. Therefore with the
increase of $\epsilon $ and $\omega $ we should expect only a
decrease of the influence of the multiple scattering on the
process of radiation \citep{BaTi86,TBHa97}.

By using the distribution function for the particles in an
electric field, the average probability over all possible
trajectories can be written as a functional integral over a Wiener
measure. The calculations have been done in \citep{BaTi86,TBHa97}
and the resulting expression for the probability of radiation of a
photon by a charged particle moving in a uniform electric field
and undergoing multiple scattering is
\begin{equation}\label{fin1}
\left\langle \frac{dI}{d\omega }\right\rangle =-\frac{e^{2}m^{2}}{\sqrt{\pi }%
\epsilon ^{2}}\left[ F_{1}\left( r,\eta ,x\right) +\left( \frac{2}{x}+\frac{%
\omega }{\epsilon }\zeta \sqrt{x}\right) F_{3}\left( r,\eta ,x\right) +\frac{%
\epsilon ^{2}+\epsilon ^{\prime 2}}{3\omega \epsilon ^{2}L\left(
\sigma _{x}\right) }F_{2}\left( r,\eta ,x\right) \right] ,
\end{equation}
where $\sigma _i$, $i=x,y$ represent the mean squares of the
angles of deviation from the direction of motion of the particle
in the average potential acquired per unit length, $\sigma
_i=d<\theta _{si}^2>/dz$, $i=x,y$, $r=\sqrt{\sigma _{y}/\sigma
_{x}}$, $\epsilon ^{\prime }=\epsilon -\omega $, $\eta
=m^{2}\omega /\epsilon \epsilon ^{\prime }\left( 2\omega \epsilon
\sigma _{x}/\epsilon ^{\prime }\right) ^{1/2}$, $x=\left(
m^{3}\omega /eE\epsilon \epsilon ^{\prime }\right) ^{2/3}$,
$L\left( \sigma _{x}\right) =\pi m^{2}/e^{2}\epsilon ^{2}\sigma
_{x}$ and
\begin{equation}
F_{1}\left( r,\eta ,x\right) =\frac{2}{\sqrt{\pi
}}\int_{0}^{\infty }e^{-\eta t}\left\{ \frac{\sin \eta
t}{2t}-f(r,t)\exp \left[ \varphi \left(
x,\eta ,t\right) \right] \sin \left[ \eta t+\varphi \left( x,\eta ,t\right) %
\right] \right\} dt,
\end{equation}
\begin{equation}
F_{2}\left( r,\eta ,x\right) =3\sqrt{2}\eta \int_{0}^{\infty
}\left[ \tanh (t)+r\tanh (rt)\right] f(r,t)\exp \left[ -\eta
t+\varphi \left( x,\eta
,t\right) \right] \cos \left[ \eta t+\varphi \left( x,\eta ,t\right) -\frac{%
\pi }{4}\right] dt,
\end{equation}
\begin{equation}
F_{3}\left( r,\eta ,x\right) =-\frac{4\eta ^{2}}{\sqrt{\pi }x^{2}}%
\int_{0}^{\infty }\tanh ^{2}(t)f(r,t)\exp \left[ -\eta t+\varphi
\left(
x,\eta ,t\right) \right] \cos \left[ \eta t+\varphi \left( x,\eta ,t\right) -%
\frac{\pi }{4}\right] ,
\end{equation}
\begin{equation}\label{fin2}
f(r,t)=\sqrt{\frac{r}{\sinh (2t)\sinh (2rt)}},\varphi \left(
x,\eta ,t\right) =\frac{2\eta ^{3}}{x^{3}}\left[ \tanh
(t)-t\right] .
\end{equation}

These equations give a complete description of the radiation
pattern emitted by a charged particle moving in a dense medium in
the presence of an electric field.

In the limit of the dominance of the electric field and symmetric
multiple scattering we have $1>x<<\eta $ and Eq. (\ref{fin1}) goes
into the expression for the intensity of the radiation in a
constant electromagnetic field, by also taking into account
quantum effects \citep{La}
\begin{equation}\label{rad1}
\frac{dI}{d\omega }=-\frac{e^{2}m^{2}\omega }{\sqrt{\pi }\epsilon
^{2}}\left[ \int_{x}^{\infty }{\rm Ai}(y)dy+\left(
\frac{2}{x}+\frac{\omega }{\epsilon }\zeta \sqrt{x}\right) {\rm
Ai}^{\prime }(x)\right] ,
\end{equation}
where $x=\left( \omega /\epsilon ^{\prime }\zeta \right) ^{2/3}$,
$\epsilon ^{\prime }=\epsilon -\omega $ and ${\rm
Ai}(y)=\int_{0}^{\infty }\cos \left( yt+t^{3}/3\right)
dt/\sqrt{\pi }$ is the Airy function. In the classical limit,
$\omega <<\epsilon ^{\prime }\approx \epsilon $, and $x\approx
\left(\omega /\omega _0\right)^{2/3}\left(m/\epsilon \right)^2$,
where $\omega _0=e\zeta /\epsilon $. In this case the second term
in the round bracket in Eq. (\ref{rad1}) is small and Eq.
(\ref{rad1}) goes into the classical formula for the
electromagnetic radiation of an electron in an external
electromagnetic field \citep{La75}.

\section{Single electron radiation, emissivity and energy flux from the
electrosphere of quark stars}

In the present Section we shall consider an example of how the
formalism described in the previous Section can be used to
describe the radiation properties of the electrosphere of the bare
strange stars.

\subsection{Structure of the electrosphere of quark stars}

For the electron distribution in the electrosphere of the bare
strange stars we use the model presented in \citet{ChHa03}, which
also takes into account the finite temperature effects and can be
solved exactly, with all physical quantities of interest (chemical
potential, electric field etc.) expressed in an exact analytical
form. The chemical equilibrium implies that the electron chemical
potential $\mu _{\infty }=-V+\mu _{e}$ is constant, where $V$ is
the electrostatic potential per unit charge and $\mu _{e}$ is the
electron's chemical potential. Since far outside the star both $V$
and $\mu _{e}$ tend to zero, it follows that $\mu _{\infty }=0$
and $\mu _{e}=V$ \citep{Al86}.

The Poisson equation for the electrostatic potential $V\left(
z,T\right) $ generated by the finite temperature electron
distribution reads \citep{Al86,Ke95}
\begin{equation}
\frac{d^{2}V}{dz^{2}}=\frac{4\alpha }{3\pi }\left[ \left(
V^{3}-V_{q}^{3}\right) +\pi ^{2}\left( V-V_{q}\right) T^{2}\right]
,z\leq 0, \label{p2}
\end{equation}
\begin{equation}
\frac{d^{2}V}{dz^{2}}=\frac{4\alpha }{3\pi }\left( V^{3}+\pi
^{2}T^{2}V\right) ,z\geq 0,  \label{3}
\end{equation}
where $T$ is the temperature of the electron layer, which can be
taken as a constant, since we assume the electrons are in
thermodynamic equilibrium with the constant temperature quark
matter. In Eqs.~(\ref{p2})-(\ref{3}), $z$ is the space coordinate
measuring height
above the quark surface, $\alpha $ is the fine structure constant and $%
V_{q}/3\pi ^{2}$ is the quark charge density inside the quark
matter. The boundary conditions for Eqs.~(\ref{p2})-(\ref{3}) are
$V\rightarrow V_{q}$ as $z\rightarrow -\infty $ and $V\rightarrow
0$ for $z\rightarrow \infty $. In the case of the zero temperature
electron distribution at the boundary $z=0$ we have the condition
$V(0)=(3/4)V_{q}$ \citep{Al86}.

The structure of the electrosphere and the corresponding radiation
processes essentially depends on the value of $V_q$, the electric
charge density inside the quark star. When the temperature of the
quark star core drops below $10^{9}$ K, the strange matter becomes
superfluid. At this temperature quarks can form colored Cooper
pairs near the Fermi surface and become superconducting. From the
BCS theory it follows that the critical temperature $T_{c}$ at
which the transition to the superconducting state takes place is
$T_{c}=\Delta /1.76$, where $\Delta $ is the pairing gap energy
\citep{Bl}. An early estimation of $\Delta $ gave $\Delta \sim
0.1-1$ MeV \cite{early}, but some recent studies considering
instanton-induced interactions between quarks estimated $\Delta
\sim 100$ MeV \citep{all1}.

Strange quark matter in the color-flavor locked (CFL) phase of
QCD, which occurs for  $\Delta \sim 100$ MeV, could be rigorously
electrically neutral, despite the unequal quark masses, even in
the presence of the electron chemical potential \cite{all1}.
Hence, for the CFL state of quark matter $V_q=0$ and no electrons
are present inside the quark star.

However, \citet{PaUs02} pointed out that for sufficiently large
$m_{s}$ the low density regime is rather expected to be in the
''2-color-flavor Superconductor'' phase in which only $u$ and $d$
quarks of two color are paired in single condensate while the ones
of the third color, and $s$ quarks of all three colors, are
unpaired. In this phase, some electrons are still present. In
other words, electrons may be absent in the core of strange stars
but present, at least, near the surface where the density is
lowest. Nevertheless, the presence of the CFL effect can reduce
the electron density at the surface and hence it can also
significantly reduces the electromagnetic emissivity of the
electrons in the surface layer. Therefore, in order to describe
the radiation properties of the electrosphere we assume that
$V_q\neq 0$.

The general solution of Eq.~(\ref{3}) is given by \citep{ChHa03}
\begin{equation}
V\left( z,T\right) =\frac{2\sqrt{2}\pi T\exp \left[
2\sqrt{\frac{\alpha \pi
}{3}}T\left( z+z_{0}\right) \right] }{\exp \left[ 4\sqrt{\frac{\alpha \pi }{3%
}}T\left( z+z_{0}\right) \right] -1},
\end{equation}
where $z_{0}$ is a constant of integration. Its value can be
obtained from the condition of the continuity of the potential
across the star's surface, requiring $V_{I}(0,T)=V\left(
0,T\right) $, where $V_{I}\left( z,T\right) $ is the value of the
electrostatic potential in the region $z\leq 0$, described by
Eq.~(\ref{p2}). Therefore
\begin{equation}
z_{0}=\frac{1}{2}\sqrt{\frac{3}{\alpha \pi }}\frac{1}{T}\ln \left[ \frac{%
\sqrt{2}\pi T}{V_{I}\left( 0,T\right) }\left( 1+\sqrt{1+\frac{%
V_{I}^{2}\left( 0,T\right) }{2\pi ^{2}T^{2}}}\right) \right] .
\end{equation}

The number density distribution $n_{e}$ of the electrons at the
quark star surface can be obtained from $n_{e}\left( z,T\right)
=V^{3}/3\pi ^{2}+VT^{2}/3$ \citep{Ke95,ChHa03}.

In the limit of zero temperature, $T\rightarrow 0$ we obtain
$V(z)=a_0/(z+b)$, where $a_0=\sqrt{3\pi /2\alpha }$ and $b$ is an
integration constant. $b$ can be determined from the boundary
condition $V(0)=(3/4)V_{q}$, which gives $b=\left(
4a_0/3V_{q}\right) $. Therefore, in this case for the electron
particle number distribution, extending several thousands fermis
above the quark matter surface, we find the expression:
$n_{e}(z)=(1/3\pi ^{2})a^{3}_{0}/ \left(z+b\right) ^{3}$.

In the absence of a crust of the quark star, the electron layer
can extend to several thousands fermis outside the star's surface.

The variation of the strength of the electric field $E$ outside
the quark star surface is given by
\begin{equation}
E\left( z,T\right) =\sqrt{\frac{2\alpha }{3\pi }}V\sqrt{V^{2}+\pi
^{2}T^{2}}.
\end{equation}

The electric field is represented as a function of the distance
from the quark star surface and for different values of the
temperature, in Fig. 1.

\subsection{Single electron radiation and emissivity of the
electrosphere of bare strange stars}

In order to study the electromagnetic radiation emission from the
electrosphere of the quark stars and the physical processes
determining it, it is necessary to find if, with respect to the
LPM effect, the electrosphere is a thin or thick medium. In
extremely thin media, neither dielectric effects nor multiple
scattering produce enough of a phase shift to cause suppression of
radiation. Suppression due to multiple scattering disappears when
the total scattering angle in the target is less than $1/\gamma $.
This happens when the thickness $l$ of the dense medium is smaller
than a critical thickness $l_c(z,T)$ (\citet{Kl99} and references
therein):
\begin{equation}
l <l_{crit}\left( z,T\right) =\left( \frac{m_e}{\epsilon _{s}}%
\right) ^{2}X_{0}=\frac{1}{4\pi }\left[ 4n\left( z,T\right)
r_{e}^{2}Z^{2}\ln \left( 184Z^{-1/3}\right) \right] ^{-1}.
\end{equation}

To describe the general properties of the electrosphere of the
quark stars with respect to the effect of the multiple scattering
on electromagnetic radiation we introduce a parameter, called
suppression factor, and defined as
\begin{equation}
s_{l}(z,T)\equiv \frac{z}{l_{crit}}=4\pi z\left[ 4n\left(
z,T\right) r_{e}^{2}Z^{2}\ln \left( 184Z^{-1/3}\right) \right].
\end{equation}

The LPM effect is important for regions for which $s_{l}(z,T)\geq
1$; for regions in the electrosphere with $s_{l}(z,T)<1$, the LPM
effect can be safely ignored. The variation of the suppression
factor $s_l$ as a function of $z$ for two sets of values of the
surface electric potential of the quark star and for different
values of the temperature is represented in Fig. 2.

As one can see from Fig. 2, the effect of the multiple scattering
is extremely important in the dense layer situated near the quark
star surface. For the electrons situated far away for the surface,
the suppression due to multiple scattering can be ignored.

In the absence of an external field electromagnetic radiation emission of electrons is strongly suppressed when the mean square multiple scattering angle over the distance $%
L_{\parallel }$, $\theta _{ms}^{2}=$ $\left( \epsilon
_{s}/\epsilon \right) ^{2}\left( L_{\parallel }/X_{0}\right) $,
where $\epsilon _{s}=m_{e}\sqrt{4\pi /\alpha }$ and $X_{0}=\left[
4n\alpha r_{e}^{2}\ln \left( 184\right) \right]
 ^{-1}$ is the radiation length, is greater than
or equal to $\theta _{\gamma }^{2}$, $\theta _{ms}^{2}\geq \theta
_{\gamma }^{2}$ \citep{Kl99,Ha03,An95}. Hence the radiation
emission differential cross section for the production of a photon
is suppressed when
\begin{equation}
\omega <\omega _{LPM}=\frac{\epsilon ^{2}}{\epsilon _{LPM}},
\end{equation}
where
\begin{equation}\label{frec}
\epsilon _{LPM}=\frac{m_{e}^{2}X_{0}\alpha }{8\pi }.
\end{equation}

For the electrosphere of the quark stars, in the calculation of
the LPM critical frequency $\omega _{LPM}$ the effect of the
external electric field has also to be included. By assuming that
the temperature of the star $T<<p_{F}=\mu _e$, with $p_{F}$ the
Fermi momentum, the Fermi distribution factors force all the
electrons to have $\left\langle \epsilon \right\rangle \approx \mu
_e\approx V$. Therefore the critical ($z$-dependent) LPM frequency
for the bremsstrahlung photon emission in the electron layer at
the surface of the quark star is given by
\begin{equation}\label{LPM}
\omega _{LPM}\left( z,T\right) \approx \frac{\mu _{e}^{2}\left( z,T\right) }{%
\epsilon _{LPM}\zeta ^{4/3}}\approx \frac{V^{2}\left( z,T\right)
}{\epsilon
_{LPM}\zeta ^{4/3}}=\frac{32r_{e}^{2}\ln (184)}{3\pi m_{e}^{2}\zeta ^{4/3}}%
V^{3}\left( z,T\right) \left[ V^{2}\left( z,T\right) +\pi
^{2}T^{2}\right] .
\end{equation}
where the parameter $\zeta =eE\epsilon /m_e^3=\gamma E/E_{cr}$
takes into account the modifications of the LPM critical frequency
due to the presence of the external electric field \citep{BaTi86}.
The LPM frequency for free electrons (in the absence of an
external field) is obtained by taking the limit $\zeta \rightarrow
1$ in Eq. (\ref{LPM}). The case $\zeta <1$ corresponds to the
classical limit, when the effects of the electric field can be
neglected. Therefore the effect of the electric field on the
electromagnetic emission in the electrosphere of the quark stars
plays an important role only for values of the electric potential
satisfying the condition
\begin{equation}
V^{2}\sqrt{V^{2}+\pi ^{2}T^{2}}>m_{e}^{3}\sqrt{3\pi /2\alpha }%
/e.
\end{equation}

Alternatively, the above condition can be formulated as a general
condition for the electric field in the electrosphere,
$E>E_{cr}/\gamma $. In the limit of small temperatures
$T\rightarrow 0$ we have $V> V_{\rm crit}$, where
\begin{equation}
V_{\rm crit}=\frac{m_{e}\left( 3\pi /2\alpha \right)
^{1/6}}{e^{1/3}}=3.40 \;{\rm MeV}.
\end{equation}

For $V<V_{\rm crit}$ the influence of the electric field of the
electrosphere on the radiation processes is negligible.

As one can see from Eq. (\ref{LPM}), all the intrinsic emission
properties of the electrosphere of the quark stars are determined
by a single parameter, the surface electric potential $V(z,T)$.
The LPM frequency is also rapidly increasing with the temperature.

We shall estimate first the numerical value of the critical
frequency for the electrons in the dense layer near the quark star's surface, for $%
z\approx 0$. In this limit $V\left( z,T\right) \rightarrow
V_{I}\left( 0,T\right) $ and by neglecting the effect of the
external electric field the LPM frequency is given by $\omega
_{LPM}^{E=0}\left( 0,T\right) =(32r_{e}^{2}\ln (184)/3\pi
m_{e}^{2})V_{I}^{3}\left( 0,T\right) \left[ V_{I}^{2}\left(
0,T\right) +\pi ^{2}T^{2}\right] $. By adopting a value of the
electric potential at the star's surface of the order of
$V_{I}\left( 0,T\right) =16$ MeV, we obtain $\omega
_{LPM}^{E=0}\left( 0,0\right) \approx 14.5$ GeV for $T=0$, $\omega
_{LPM}^{E=0}\left( 0,1\right) \approx 15$ GeV for $T=1$ MeV and
$\omega _{LPM}^{E=0}\left( 0,10\right) \approx 70$ GeV for $T=10$
MeV. However, by adopting for the electric potential at the star's
surface a value of the order of $V_{I}(0,T)=1$ MeV, the value of
the LPM frequency is only $\omega _{LPM}^{E=0}\left( 0,0\right)
\approx 0.013$ MeV
for $T=0$, $\omega _{LPM}^{E=0}\left( 0,1\right) \approx 0.15$ MeV and $%
\omega _{LPM}^{E=0}\left( 0,10\right) \approx 13.6$ MeV for $T=10$
MeV.

The presence of the electric field in the electrosphere of the
quark stars strongly suppresses the LPM effect. For $V_{I}=16$ MeV
and for $T=0$ the value of the electric field is $E=10$ MeV$^{2}$
and $\zeta ^{4/3}=488$. The corresponding value of the LPM
frequency $\omega _{LPM}^{E\neq 0}\left(
0,0\right) \approx 29.61$ MeV. For $T=1$ MeV, $\zeta ^{4/3}=500.5$ and $%
\omega _{LPM}^{E\neq 0}\left( 0,1\right) \approx 30$ MeV, while
for $T=10$ MeV, $\omega _{LPM}^{E\neq 0}\left( 0,10\right) \approx
29.61$ MeV, $\zeta ^{4/3}=1400$ and $\omega _{LPM}^{E\neq 0}\left(
0,1\right) \approx 50.15$ MeV. For $V_{I}=1$ MeV, since
$V_I<V_{\rm crit}$, the values of the LPM frequency are the same
as in the case of the free electrons, $\left.\omega _{LPM}^{E\neq
0}\left( 0,T\right)\right|_{V_I=1{\rm MeV}} \approx \left.\omega
_{LPM}^{E=0}\left( 0,T\right)\right|_{V_I=1 {\rm MeV}}$.

Emission of the electromagnetic radiation with frequency smaller
than $\omega _{LPM}$ from the electron layer of the quark star
surface is suppressed, due to the effect of multiple scattering.
As expected, the LPM frequency decreases with the density of the
electrons at the star's surface, and for $z\geq 500$ fm the effect
of the multiple scattering can be neglected.

The variation of the LPM frequency as a function of $z$ in the
presence and in the absence of an electric field is represented,
for two different values of the temperature, in Fig. 3.

The mean value of the LPM frequency, $\left\langle \omega
_{LPM}\right\rangle $, can be easily obtained in the low temperature limit, $%
T\rightarrow 0$. In this case, by neglecting the effect of the
external electric field and by taking into account that the effect
of multiple scattering has a significant effect on the electron
radiation only for a distance of the order $z_{crit}$, we obtain,
\begin{equation}
\left\langle \omega _{LPM}\right\rangle_{E=0}
=\frac{27}{32}\sqrt{\frac{3}{2\pi
\alpha }}\frac{r_{e}^{2}\ln (184)}{m_{e}^{2}z_{crit}}V_{q}^{4}%
\left[ 1-\left( 1+\frac{\sqrt{6\alpha }z_{crit}V_{q}}{4\sqrt{\pi
}}\right) ^{-4}\right] .
\end{equation}

The value of $z_{crit}$ can be obtained as a solution of the
algebraic equation $z_{crit}n\left( z_{crit},T\right) =1/16\pi
r_{e}^{2}Z^{2}\ln \left( 184Z^{-1/3}\right)$. The mean frequency
of the emitted radiation essentially depends on the value of the
electric potential at the quark star surface $V_{q}$ and of the
value of the electric field. The mean value of the electron energy
in the dense layer of the electrosphere, $\left\langle \epsilon
\right\rangle =\left( 1/z_{crit}\right)
\int_{0}^{z_{crit}}V(z)dz$, can be represented as
\begin{equation}
\left\langle \epsilon \right\rangle =\sqrt{\frac{3\pi }{2\alpha }}\frac{1}{%
z_{crit}}\ln \left( 1+\frac{\sqrt{6\alpha }z_{crit}V_{q}}{4\sqrt{\pi }}%
\right) ,T=0,
\end{equation}
\begin{equation}
\left\langle \epsilon \right\rangle =\sqrt{\frac{3\pi }{2\alpha }}\frac{1}{%
z_{crit}}\ln \frac{\left( e^{2\sqrt{\frac{\alpha \pi
}{3}}Tz_{0}}+1\right) \left( e^{2\sqrt{\frac{\alpha \pi
}{3}}T\left( z_{0}+z_{crit}\right) }-1\right) }{\left(
e^{2\sqrt{\frac{\alpha \pi }{3}}Tz_{0}}-1\right) \left(
e^{2\sqrt{\frac{\alpha \pi }{3}}T\left( z_{0}+z_{crit}\right) }+1\right) }%
,T\neq 0.
\end{equation}

For the mean value of the electric field in the region $0\leq
z\leq z_{crit}$ we obtain
\begin{equation}
\left\langle E\right\rangle =\frac{3V_{q}}{4z_{crit}}\left[ 1-\left( 1+\frac{%
\sqrt{6\alpha }z_{crit}V_{q}}{4\sqrt{\pi }}\right) ^{-1}\right] ,
T=0.
\end{equation}

The mean value of the parameter $\zeta =\gamma E/E_{cr}=eE\epsilon /m_{e}^{3}=\sqrt{%
2\alpha /3\pi }eV^{3}/m_{e}^{3}$, describing the effect of the
electric field on the radiation processes in the electrosphere of
the quark stars is given by
\begin{equation}
\left\langle \zeta \right\rangle =\frac{27\pi }{64\alpha m_eE_{cr}}\sqrt{\frac{3\pi }{%
2\alpha }}\frac{V_{q}^{2}}{z_{crit}}\left[ 1-\left( 1+%
\frac{\sqrt{6\alpha }z_{crit}V_{q}}{4\sqrt{\pi }}\right)
^{-2}\right] ,T=0.
\end{equation}

By including the effect of the external electric field, the mean
value of the LPM frequency can be expressed as
\begin{equation}
\left\langle \omega _{LPM}\right\rangle_{E\neq 0} =\frac{32\left(
3\pi \right)
^{1/6}r_{e}^{2}E_{cr}\ln (184)}{\left( 2\alpha \right) ^{7/6}e^{1/3}z_{crit}%
}\ln \left( 1+\frac{\sqrt{6\alpha }z_{crit}V_{q}}{4\sqrt{\pi
}}\right) ,T=0.
\end{equation}

In the absence of the electric field the dependence of the LPM
critical frequency on the surface potential of the quark star is
$\left\langle \omega _{LPM}\right\rangle \sim V_{q}^{4}$, while in
the presence of the electric field we have a logarithmic
dependence on the quark star potential. The dependence of the mean
LPM critical frequency $\left\langle \omega _{LPM}\right\rangle $
on the quark star surface potential $V_q$, in the low temperature
limit and with and without the presence of an external electric
field is represented in Fig. 4.

In order to calculate the spectral distribution of the intensity
of the electromagnetic radiation emitted by an electron in the
dense layer of the electrosphere of a quark star, by taking into
account the effect of the multiple collisions and of the electric
field, we use Eqs. (\ref{fin1}-\ref{fin2}). The variation of the
probability of emission of a photon by an electron moving in the
electrosphere of the quark star is represented, for two sets of
values of the quark star surface electric potential, in Fig. 5.

The mean value of the total intensity of the radiation,
$\left\langle I(\omega )\right\rangle $ is represented in Fig. 6.

An important parameter describing the radiation properties of an
electron gas is the plasma frequency $\omega _p$, defined as
$\omega _{p}=\sqrt{4\pi en_{e}(z,T)/\mu _{e}(z,T)}$, and which is
related to the medium permittivity $\varepsilon \left( \omega
\right) $ by $\varepsilon \left( \omega \right) \approx 1-\omega
_{p}^{2}/\omega ^{2}$, $\omega >>\omega _{p}$ \citep{Ja75}. The
polarization of the electrosphere of the quark stars also strongly
influences the radiation processes, and photons with frequency
$\omega <\omega _{p}$ cannot escape from the electrosphere. The
variation of the plasma frequency of the quark star electrosphere
is represented, as a function of distance, in Figs. 7.

The variation of the emissivity of the electrosphere of the quark
star, by fully taking into the effect of the electric field on the
suppression of the radiation due to the multiple scattering
effects, is presented in Figs. 8.

The total energy flux of the electrosphere is obtained as
$F_{\gamma }^{(ee)}=(1/\pi )\int_{0}^{\infty }\varepsilon _{\gamma
}^{(ee)}dz $. The variation of the energy flux as a function of
the temperature of the quark star surface is presented, for
different values of the surface electric potential $V_{I}$, in
Fig. 9.

\section{Discussions and final remarks}

Since the proposal of strange quark stars, much effort has been
devoted to find major observational properties that distinguish
strange stars from neutron stars. One of the important difference
between them is the stellar radius. Strange stars could have a
radius significantly smaller than that of neutron stars. The
observed thermal X-ray properties of compact objects may be a
possible probe to determine the stellar radius. In the present
paper we have shown that the high-density degenerate electron
layer at the surface of a bare strange star could play an
important role in the electromagnetic radiation emission from the
star. The bremsstrahlung emissivity of the electrosphere is
generally a function of the electron density, temperature and of
the electrostatic potential at the surface of the star. The
bremsstrahlung emission spectrum is very different from the black
body spectrum. For hot strange stars, the energy emission from the
electrosphere could exceed the intensity of the black body
radiation. Even in the low temperature limit the photon emission
from the electron layer exceeds the bremsstrahlung emissivity of
the quark-gluon plasma. This makes the process of
electron-electron bremsstrahlung of considerable importance for
establishing the correct observational signatures of quark stars.
Since the photon emission from the strange star's surface is
dominated by the bremsstrahlung emission from the electron layer,
it is necessary to include this effect in determining the radius
of the star.

One of the important features of the bremsstrahlung radiation from
a single interaction is the independence of $dI/d\omega $ on
$\omega $ \citep{La75}. But, as one can see from Eq. (\ref{est}),
in the case of the strongly suppressed radiation of a particle in
a dense medium we have $dI/d\omega \propto \omega ^{1/2}$.
Therefore, due to the multiple scattering effects, the spectrum of
the radiation of the dense electrosphere differs in a qualitative
way of the standard bremsstrahlung. On the other hand the electric
field of the electrosphere decreases the LPM critical frequency,
thus leading to an increase of the electromagnetic energy flux
from the bare strange star surface.

The results on the bremsstrahlung radiation of the electrosphere
obtained in the present paper have been obtained under two main
assumptions. Firstly, we have considered the equations for the
radiation emission process derived for the case of a constant
electric field. The electric field in the electrosphere is not
constant and the in order to take in account the variations of $E$
and of the energies of the electrons we have adopted the mean
values of these quantities. This approximation is of course not
generally valid for all layers of the electrosphere, but we still
expect that it can provide a general description of the radiation
pattern of the electron layer. Secondly, in describing the LPM
effect, we have adopted for the electrons the formulas and
equations describing multiple Coulomb scattering in amorphous
media, where the main contribution to scattering comes from fixed
centers, such as the nuclei or the crystalline lattice. To apply
this well-established formalism to the case of the electrosphere
of the strange stars we have assumed that the multiple scattering
processes between electrons can be described by taking $Z=1$ in
the equations describing multiple collisions and radiation
suppression in amorphous media. In other words, we approximate the
multiple collision and scattering processes between electrons in
the dense electrosphere as scattering of an electron on a fixed
center. The differences between these physical processes could
affect the numerical estimations for the suppression factor,
critical LPM frequency and radiation spectrum obtained in the
present paper. However, we expect that the results obtained in the
present paper could give a good qualitative description of the
radiation emission processes from the electrosphere.

The radiation properties of the electrosphere and the LPM
suppression effect essentially depend on the numerical value of
the electrostatic potential at the quark star surface $V_I(0,T)$,
which at low temperatures is related to the quark electric
potential $V_q$ in the bulk. It is now commonly accepted that
because of the attractive interaction between quarks in some
specific channels, the ground state of the strange quark matter is
a color superconductor, which, at asymptotic densities, is in the
color-flavor locked (CFL) phase, in which quarks of all three
flavors and three colors are paired in a single condensate. The
CFL phase in the bulk consists of equal numbers of $u$, $d$ and
$s$ quarks and is electrically neutral in the absence of any
electrons, that is, in this state $V_q=0$. Therefore, for strange
stars in the CFL phase there could be no surface electron layer,
as suggested by \citet{LuHo03}.

On the other hand the density of quark states near the surface of
a strange star differs from the density of quark states in the
bulk \citep{Ma00b,Ma01}. This results in a sharp increase of quark
charge density at the star's surface. Because of the surface
depletion of the $s$ quarks, a thin charged layer forms at the
surface of the star. Due to this surface effect the electric field
and the density of electrons could increase by an order of
magnitude \citep{Us04,UsHaCh04}. The inclusion of the surface or
other effects in the electrosphere models would also require the
reconsideration of the radiation emission processes by taking into
account their influence on the radiation suppression mechanisms
and on the critical LPM frequency.

\acknowledgments

This work is supported by a RGC grant of the Hong Kong Government.
We are grateful to the anonymous referee for comments which
significantly improved the manuscript.

\clearpage



\begin{figure}
\plotone{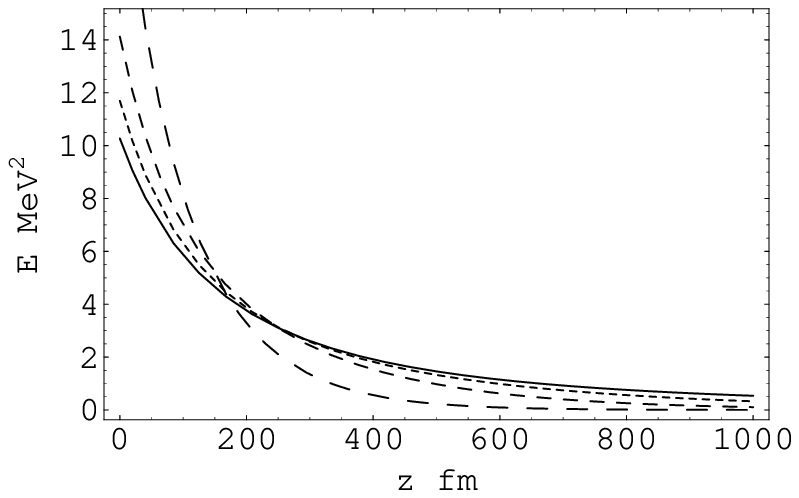} \caption{The electric field $E$ (in units of
MeV$^2$), as a function of the distance $z$ (fm), for different
values of the temperature: $T=1$ MeV (solid curve), $T=3$ MeV
(dotted curve), $T=5$ MeV (dashed curve) and $T=10$ MeV (long
dashed curve). In all cases $V_{I}(0,T)=16$ MeV. \label{FIG1}}
\end{figure}

\begin{figure}[h]
\plottwo{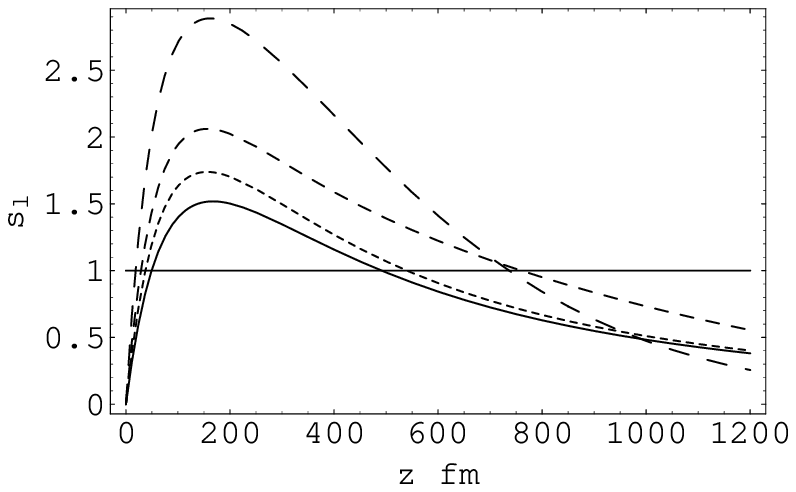}{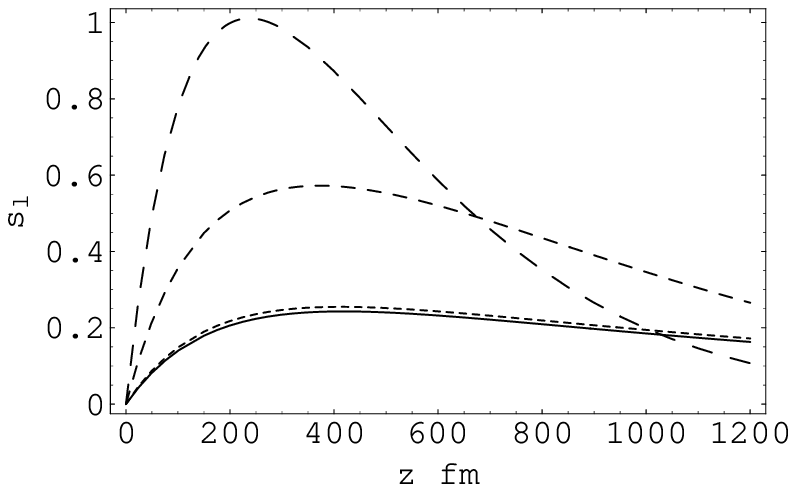} \caption{Variation of the suppression
factor $s_l$ for (a) $V_q=20$ MeV ($T=0$) and $V_{I}(0,T)=16$ MeV
($T\neq 0$) and (b) $V_q=8$ MeV ($T=0$) and $V_{I}(0,T)=6$ MeV
($T\neq 0$), as a function of the distance $z$ (fm) for different
values of the temperature: $T=0$ (solid curve), $T=0.5$ MeV
(dotted curve), $T=2.5$ MeV (dashed curve) and $T=4.5$ MeV (long
dashed curve).}\label{Fig2}
\end{figure}

\begin{figure}[h]
\plotone{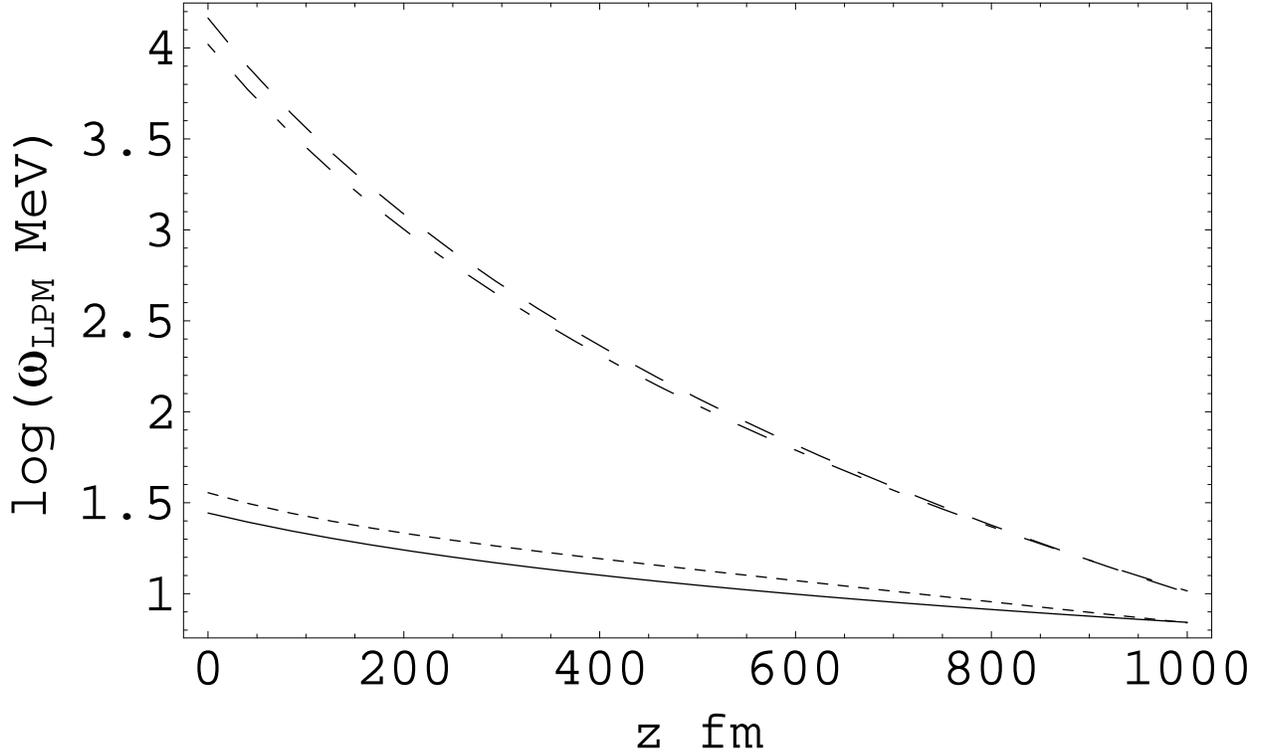} \caption{The LPM frequency $\omega _{LPM}$ (in a
logarithmic scale) of the radiation emitted in the electrosphere
of the quark stars in the presence and in the absence of the
electric field, for different values of the temperature: $T=0$ and
$E\neq 0$ (solid curve), $T=1.5$ MeV and $E\neq 0$ (short dashed
curve), $T=0$ and $E=0$ (dashed-dotted curve) and $T=1.5$ MeV and
$E=0$ (long dashed curve). For the electric potential of the quark
star surface we have considered the values $V_q=20$ MeV ($T=0$)
and $V_{I}(0,T)=16$ MeV ($T\neq 0$).}\label{Fig3}
\end{figure}

\begin{figure}[h]
\plotone{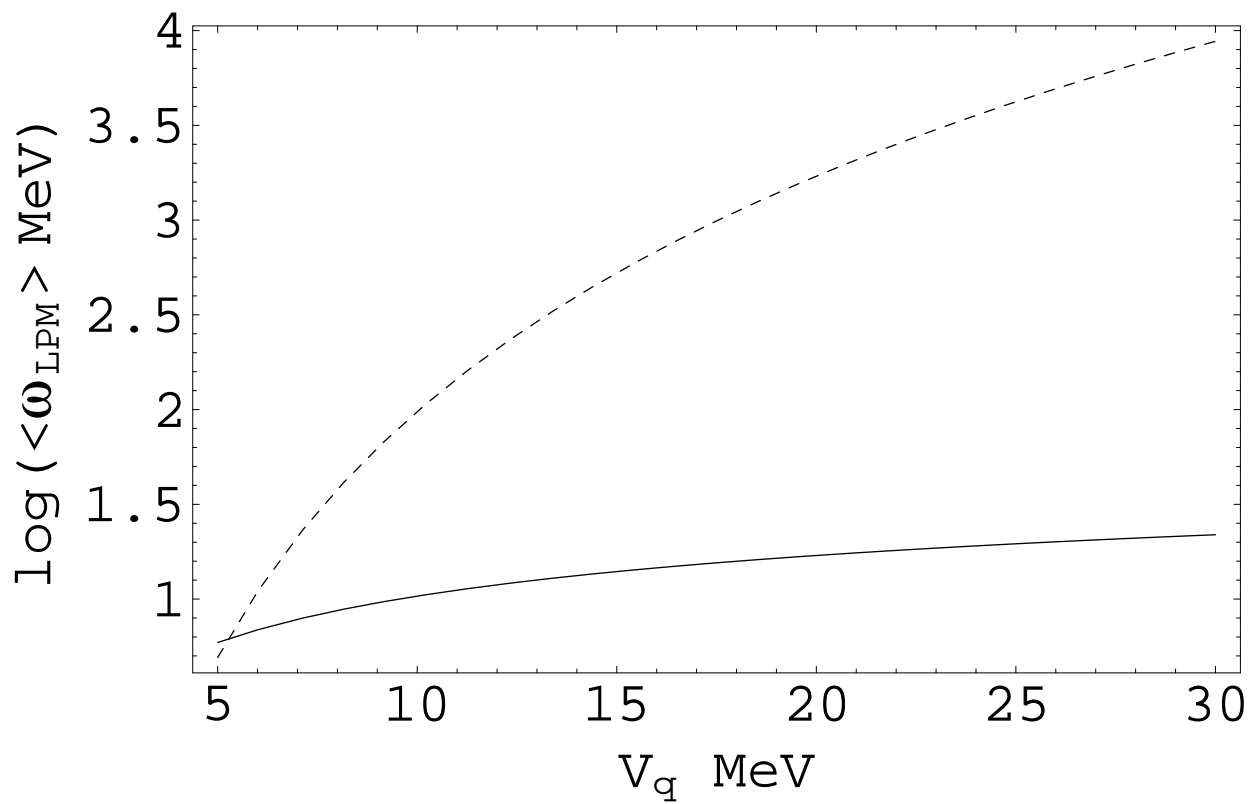} \caption{Mean LPM frequency $\left\langle \omega
_{LPM}\right\rangle$ (in a logarithmic scale) of the radiation
emitted in the electrosphere of the cold quark stars ($T<40$ MeV),
as a function of the quark star surface electrostatic potential
$V_q$, in the presence of a constant external electric field,
$E\neq 0$ (solid curve) and with the effects of the external field
neglected, $E=0$ (dotted curve).}\label{Fig4}
\end{figure}

\begin{figure}[h]
\plottwo{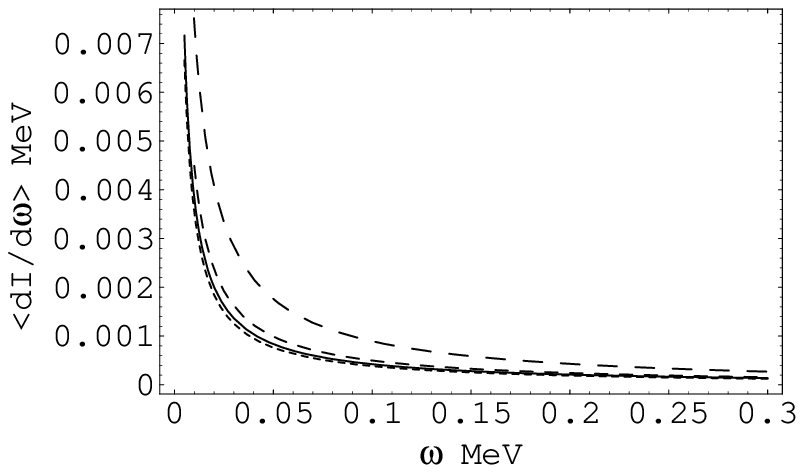}{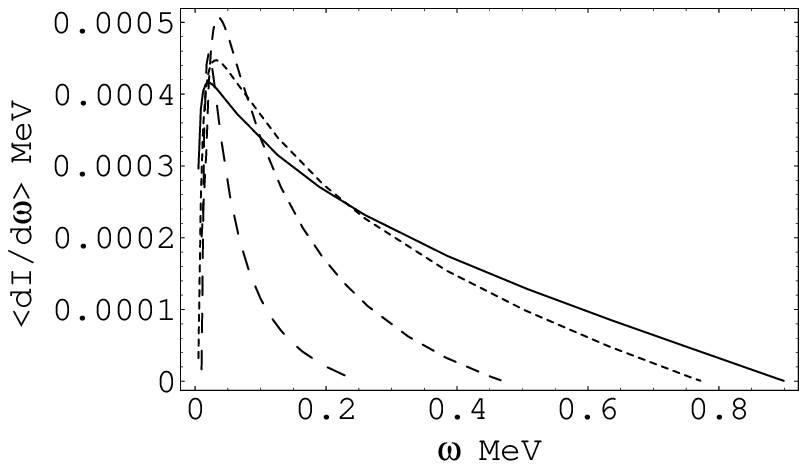} \caption{Frequency distribution of the
mean intensity $\left\langle dI/d\omega \right\rangle$ of the
radiation emitted by an electron moving in the exterior electric
field of the quark star and undergoing multiple scattering for (a)
$V_I(0,T)=16$ MeV ($T\neq 0$) and (b) $V_I(0,T)=1$ MeV ($T\neq
0$), as a function of the frequency $\omega $ of the radiation,
for different values of the temperature: $T=0.5$ MeV (solid
curve), $T=1$ MeV (dotted curve), $T=2.5$ MeV (dashed curve) and
$T=4.5$ MeV (long dashed curve). For the energy of the electron
and the electric field in the electrosphere we have adopted the
mean values corresponding to a given temperature, while $\sigma
_x=0.1$ MeV and $r=2$.}\label{Fig5}
\end{figure}

\begin{figure}
\plotone{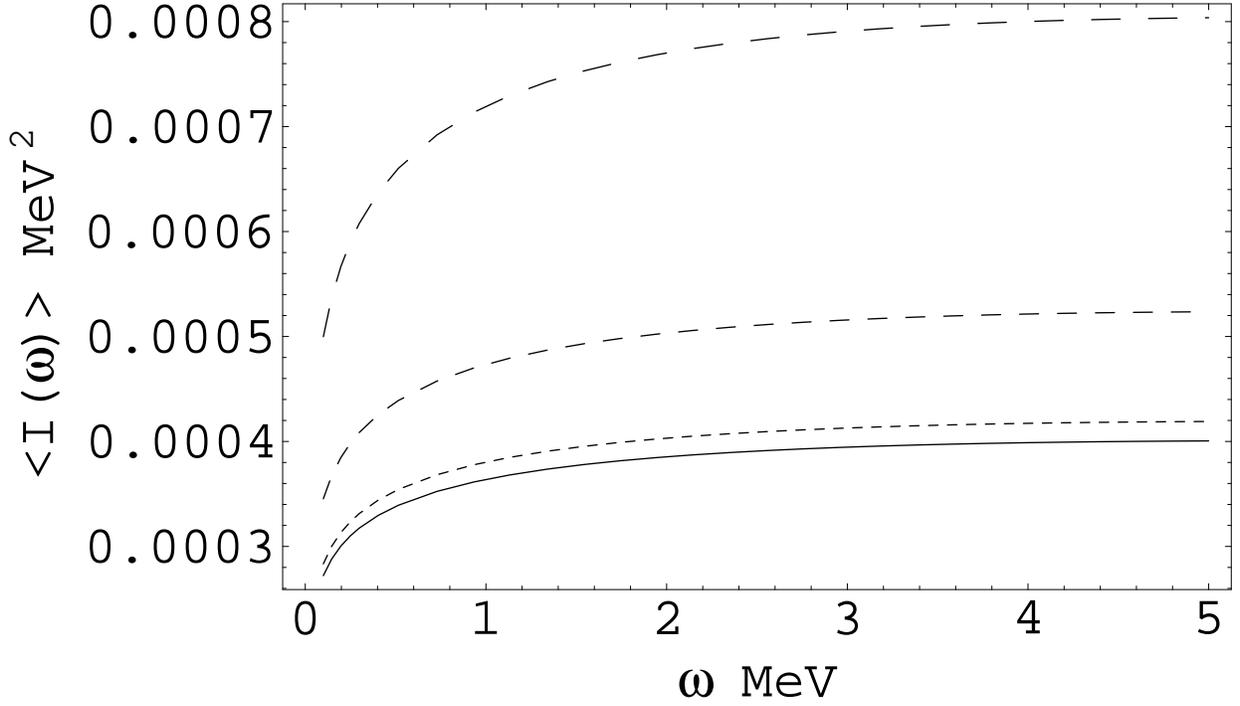}
\caption{Mean intensity $\left\langle I (\omega ) \right\rangle$
of the radiation emitted by an electron moving in the electric
field of the quark star, and undergoing multiple scattering, as a
function of the frequency $\omega $ of the radiation, for
different values of the temperature: $T=0$ MeV (solid curve),
$T=1.5$ MeV (dotted curve), $T=2.5$ MeV (dashed curve) and $T=4.5$
MeV (long dashed curve). For the value of the quark surface
electric potential we have chosen $V_q=20 $ MeV ($T=0$) and
$V_{I}(0,T)=16$ MeV ($T\neq 0$), while $\sigma _x=0.1$ MeV and
$r=2$. For the energy of the electron and the electric field in
the electrosphere we have adopted the mean values corresponding to
a given temperature.} \label{FIG6}
\end{figure}

\begin{figure}[h]
\plottwo{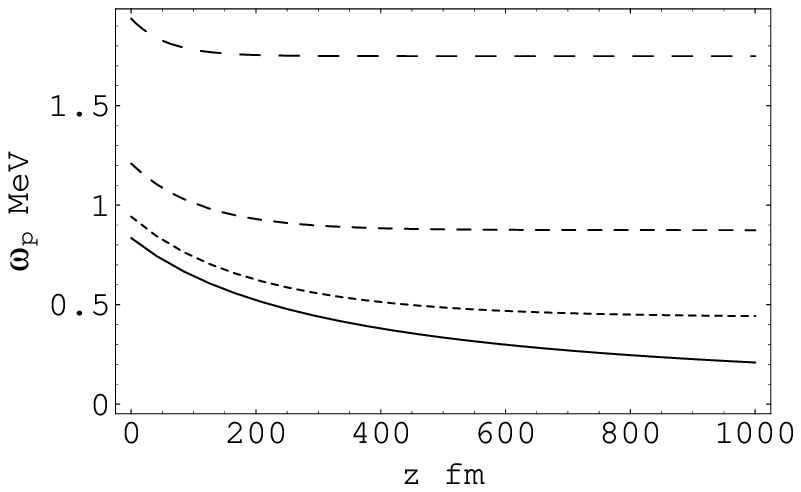}{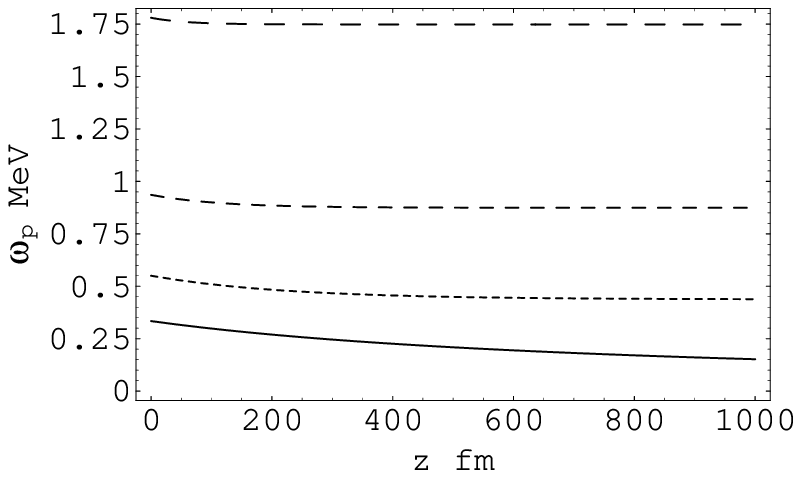} \caption{Plasma frequency $\omega _p$
of the electrosphere of the quark stars as a function of the
distance $z$ for (a) $V_q=20$ MeV ($T=0$) and $V_I(0,T)=16$ MeV
($T\neq 0$) and (b) $V_q=8$ MeV ($T=0$) and $V_I(0,T)=6$ MeV
($T\neq 0$), for different values of the temperature: $T=0$ MeV
(solid curve), $T=2.5$ MeV (dotted curve), $T=5$ MeV (dashed
curve) and $T=10$ MeV (long dashed curve).}\label{Fig7}
\end{figure}

\begin{figure}[h]
\plottwo{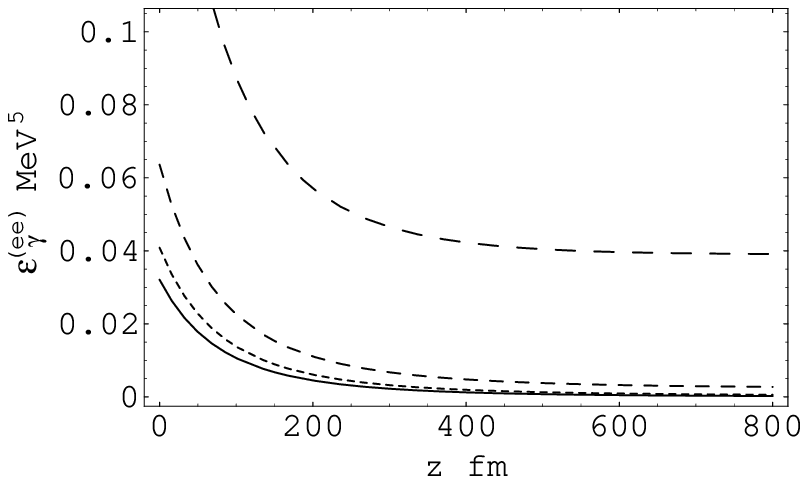}{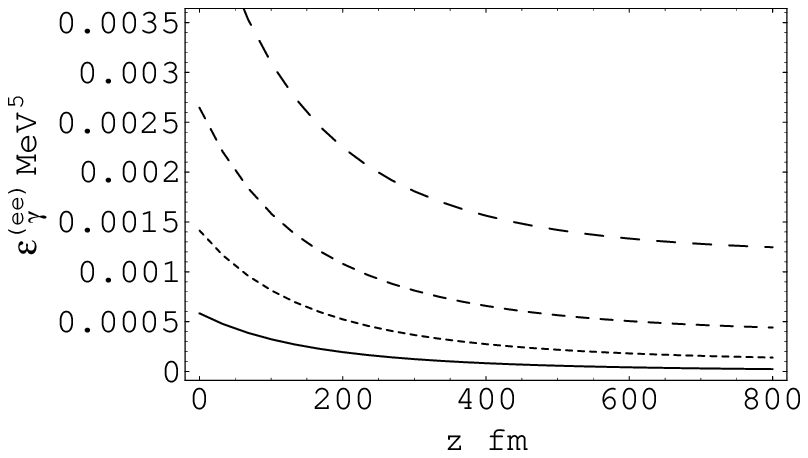} \caption{Emissivity $\varepsilon
_\gamma ^{(ee)}(z,T)$ of the electrosphere of the quark stars as a
function of the radial distance $z$ for (a) $V_I=16$ MeV ($T\neq
0$) and (b) $V_I(0,T)=8$ MeV ($T\neq 0$), for different values of
the temperature: $T=0.5$ MeV (solid curve), $T=1.5$ MeV (dotted
curve), $T=2.5$ MeV (dashed curve) and $T=4.5$ MeV (long dashed
curve). For the energy of the electron and the electric field in
the electrosphere we have adopted the mean values corresponding to
a given temperature, while $\sigma _x=0.1$ MeV and
$r=2$.}\label{Fig8}
\end{figure}

\begin{figure}
\plotone{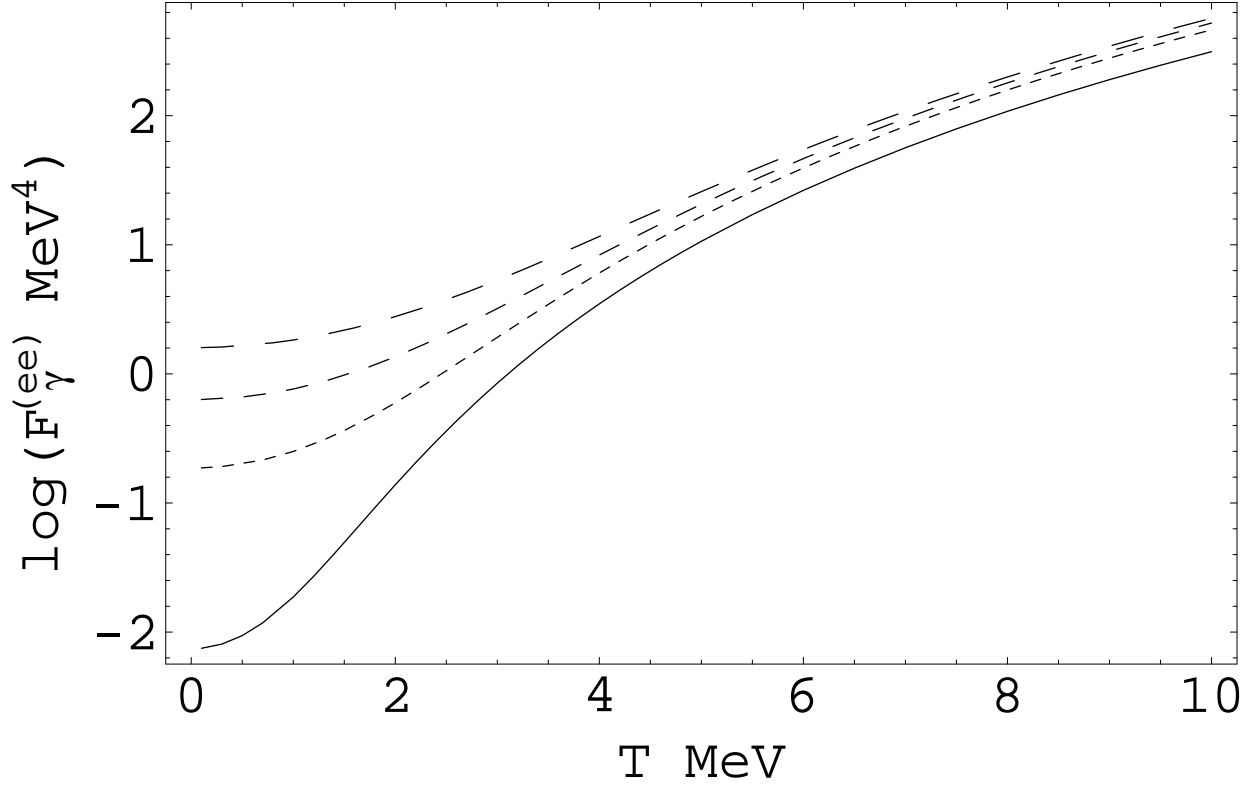}
\caption{Variation, as a function of the temperature $T$, of the
energy flux $F_{\gamma }^{(ee)}$ emitted by the electrosphere of a
quark star (in a logarithmic scale), for different values of the
quark star surface electric potential $V_I$: $V_I=8$ MeV (solid
curve), $V_I=12$ MeV (dotted curve), $V_I=16$ MeV (dashed curve)
and $V_I=20$ MeV (long dashed curve). For the energy of the
electron and the electric field in the electrosphere we have
adopted the mean values corresponding to a given temperature,
while $\sigma _x=0.1$ MeV and $r=2$.} \label{FIG9}
\end{figure}


\end{document}